\documentclass[twocolumn,showpacs]{revtex4}
\usepackage{amssymb}
\usepackage{amsmath}
\usepackage{dcolumn}
\usepackage{graphicx}
\usepackage{bm}

\begin{document}

\title{Spin-quadrupole ordering of spin-3/2 ultracold fermionic atoms in
optical lattices in the one-band Hubbard model}
\author{Hong-Hao Tu,$^{1}$ Guang-Ming Zhang,$^{1}$ and Lu Yu$^{2}$}
\affiliation{$^{1}$Department of Physics, Tsinghua University, Beijing 100084, China\\
$^{2}$Institute of Theoretical Physics, Chinese Academy of Sciences, Beijing
100080, China}
\date{\today}

\begin{abstract}
Based on a generalized one-band Hubbard model, we study magnetic
properties of Mott insulating states for ultracold
spin-$\frac{3}{2}$ fermionic atoms in optical lattices. When the
\textit{s}-wave scattering lengths for the total spin $S=2,0$
satisfy conditions $a_{2}>a_{0}>0$, we apply a functional integral
approach to the half filled case, where the spin-quadrupole
fluctuations dominate. On a 2D square lattice, the saddle point
solution yields a staggered spin-quadrupole ordering at zero
temperature with symmetry breaking from SO(5) to SO(4). Both spin
and spin-quadrupole static structure factors are calculated,
displaying highly anisotropic spin antiferromagnetic fluctuations
and antiferroquadrupole long-range correlations, respectively.
When Gaussian fluctuations around the saddle point are taken into
account, spin-quadrupole density waves with a linear dispersion
are derived. Compared with the spin density waves in the half
filled spin-$\frac{1}{2}$ Hubbard model, the quadrupole density
wave velocity is saturated in the strong-coupling limit, and there
are no transverse spin-quadrupole mode couplings, as required by
the SO(4) invariance of the effective action. Finally, in the
strong-coupling limit of the model Hamiltonian, we derive the
effective hyperfine spin-exchange interactions for the Mott
insulating phases in the quarter filled and half filled cases,
respectively.
\end{abstract}

\maketitle

\section{Introduction}

Ultracold atoms in optical lattices provide us with an ideal playground to
study various interesting quantum phenomena and to explore novel many-body
states with no counterparts in traditional solid state systems \cite%
{Lewenstein-2006}. For alkali atoms, their hyperfine spin $F$ is given by a
combination of the nuclear spin $I$ and the electron spin $S=\frac{1}{2}$,
which has $2F+1$ manifold magnetic states. In magnetic traps, these $2F+1$
components are split, while in optical traps these hyperfine spin degrees of
freedom are degenerate and strong quantum fluctuations are displayed.
Depending on the value of the hyperfine spin $F$, one classifies ultracold
atoms as bosons with integers and fermions with half integers.

It has been well-established that ultracold bosonic atoms in optical
lattices can exhibit a variety of novel phenomena, such as spinor Bose
condensation \cite{Ketterle-1998,Ho-1998,Machida-1998,Demler-Zhou-2002},
coherent spin dynamics \cite{Bloch-2005,You-2005} and superfluid-Mott
insulator quantum phase transition \cite{Bloch-2002,Kurihara-2005}.

Meanwhile a degenerate $F=\frac{3}{2}$ Fermi gas can be obtained by cooling
alkali atoms $^{132}$Cs, as well as alkaline-earth atoms $^{9}$Be, $^{135}$%
Ba, and $^{137}$Ba. These high spin fermionic atoms in optical traps may
lead to peculiar many-body ground states and exotic collective excitations,
rarely appearing in interacting electron systems \cite{Ho-1999,CWu-2003}.
For example, the Cooper pair structures in such high spin fermions are
enriched: a kind of quintet pairing can be formed as an SO(5) polar
condensate \cite{CWu-Hu-2005}. In one dimension, the competing SO(5)
superfluid order has been investigated based on the bosonization technique
\cite{CWu-2005, Lecheminant-2005}, and an exact Bethe-ansatz method \cite%
{Tsvelik-2006} has also been used to describe the corresponding low-energy
states. Furthermore, a spin-3/2 ladder model has been proposed with
spontaneous plaquette ground state \cite{YPWang-2005}.

Out of these intensive studies, Mott insulating states of ultracold atoms in
optical lattices are of particular interest. With a fixed number of atoms on
each lattice site, the quantum fluctuations of the hyperfine spin degrees of
freedom may lead to magnetic multipolar long-range order. Most previous
studies focused on ultracold bosons with hyperfine spin $F=1,2$.
Adiabatically increasing optical lattice trap depth, these systems go
through a quantum superfluid-Mott insulator phase transition. The atomic
virtual tunneling processes can result in effective hyperfine spin-exchange
interactions between the nearest neighbor sites. In particular, an
insulating spin-quadrupole or spin nematic ordering has been proposed for
spin-1 bosons in optical lattices \cite%
{Demler-Zhou-2002,Demler-2003,Zhou-2004}. This is a quantum analogue of
liquid crystal states, where the spin SU(2) rotational symmetry is broken
while the time reversal symmetry is preserved \cite%
{Andreev-1984,Barzykin-1991}. However, direct experimental observation of
such a spin nematic ordering in correlated electron systems by conventional
probes is rather difficult \cite{Demler-2005}. In this paper, we will
demonstrate that such a novel quantum magnetic state also exists in the
ultracold spin-$\frac{3}{2}$ fermionic atoms in optical lattices, which may
provide new opportunities for experimental investigations.

In Sec. II, we will introduce the general spin-$\frac{3}{2}$ Hubbard model
based on the microscopic \textit{s}-wave atom-atom interactions. In Sec. III
a functional integral approach is applied to the half filled generalized
Hubbard model when the \textit{s}-wave scattering lengths for the total spin
$S=2,0$ satisfy conditions $a_{2}>a_{0}>0$. On a square lattice, the
saddle-point solution gives rise to a staggered spin-quadrupole (nematic)
ordered state with symmetry breaking from SO(5) to SO(4), and the
corresponding spin and spin-quadrupole correlation functions will be
evaluated. Moreover, by taking into account the Gaussian fluctuations around
the saddle point, the spin-quadrupole (nematic) density waves are derived
and the corresponding density wave velocity and stiffness are calculated. In
Sec. IV, using the second-order perturbation theory in the strong-coupling
limit of the model Hamiltonian, we will derive the effective hyperfine
spin-exchange interactions for the Mott insulating phases in the quarter
filled and half filled cases, respectively. Finally, a summary is presented
in Sec. V.

\section{Formulation of spin-$\frac{3}{2}$ one-band Hubbard model}

In order to present our results in a self-contained way and to introduce
notations, we first rewrite the fundamental interactions between two
hyperfine spin $F=\frac{3}{2}$ fermion atoms with a contact potential \cite%
{Ho-1999,CWu-2003}. For the low energy states of fermions, it is convenient
to consider only the \textit{s}-wave scattering,
\begin{equation}
V(\mathbf{r}_{1}-\mathbf{r}_{2})=\frac{4\pi \hbar ^{2}}{m}\delta (\mathbf{r}%
_{1}-\mathbf{r}_{2})(a_{0}\mathcal{P}_{0}+a_{2}\mathcal{P}_{2}),
\end{equation}%
where $\mathcal{P}_{S}$ projects the pair of atoms into states with total
spin $S=0$, $2$ and $a_{S}$ is the \textit{s}-wave scattering length in the
spin-$S$ channel. Due to antisymmetry of the wave function, the \textit{s}%
-wave scattering of identical fermions in channels $S=1,3$ is not allowed.
Using the relations%
\begin{equation*}
\mathcal{P}_{0}+\mathcal{P}_{2}=1,\text{ }\mathbf{S}_{1}\mathbf{\cdot S}%
_{2}=\lambda _{0}\mathcal{P}_{0}+\lambda _{2}\mathcal{P}_{2},
\end{equation*}%
with $\lambda _{0}=-15/4$ and $\lambda _{2}=-3/4$, the interaction can be
written in terms of spin operators%
\begin{equation}
V(\mathbf{r}_{1}-\mathbf{r}_{2})=\delta (\mathbf{r}_{1}-\mathbf{r}%
_{2})\left( g_{0}+g_{2}\mathbf{S}_{1}\mathbf{\cdot S}_{2}\right) ,
\end{equation}%
where $g_{0}=\pi \hbar ^{2}(5a_{2}-a_{0})/m$ and $g_{2}=4\pi \hbar
^{2}(a_{2}-a_{0})/(3m)$.

By introducing fermionic creation operators $\psi _{i\alpha }^{\dag }$ for
states in the lowest Bloch band on site $i$ with spin components $\alpha
=3/2,1/2,-1/2,-3/2$, a generalized one-band Hubbard model can be written as:%
\begin{eqnarray}
H &=&-t\sum_{\langle ij\rangle ,\alpha }(\psi _{i\alpha }^{\dag }\psi
_{j\alpha }+\text{H.c.})+\frac{c_{0}}{2}\sum_{i}N_{i}(N_{i}-1)  \notag \\
&&+\frac{c_{2}}{2}\sum_{i}\left( \mathbf{S}_{i}^{2}-\frac{15}{4}N_{i}\right)
-\mu \sum_{i}N_{i},  \label{Hamil}
\end{eqnarray}%
where the coupling parameters $c_{0}$ and $c_{2}$ are proportional to $g_{0}$
and $g_{2}$, respectively, while their explicit expressions depend on the
potential of optical traps and the atomic recoil energy. The total number of
atoms on site $i$ is
\begin{equation*}
N_{i}=\sum_{\alpha }\psi _{i\alpha }^{\dag }\psi _{i\alpha },
\end{equation*}%
and the total spin on site $i$ is defined by
\begin{equation*}
\mathbf{S}_{i}=\sum_{\alpha \beta }\psi _{i\alpha }^{\dag }\mathbf{S}%
_{\alpha \beta }\psi _{i\beta },
\end{equation*}%
with the spin-3/2 matrices%
\begin{eqnarray}
S^{x} &=&\left(
\begin{array}{cccc}
0 & \frac{\sqrt{3}}{2} & 0 & 0 \\
\frac{\sqrt{3}}{2} & 0 & 1 & 0 \\
0 & 1 & 0 & \frac{\sqrt{3}}{2} \\
0 & 0 & \frac{\sqrt{3}}{2} & 0%
\end{array}%
\right) ,  \notag \\
S^{y} &=&\left(
\begin{array}{cccc}
0 & -i\frac{\sqrt{3}}{2} & 0 & 0 \\
i\frac{\sqrt{3}}{2} & 0 & -i & 0 \\
0 & i & 0 & -i\frac{\sqrt{3}}{2} \\
0 & 0 & i\frac{\sqrt{3}}{2} & 0%
\end{array}%
\right) ,  \notag \\
S^{z} &=&\left(
\begin{array}{cccc}
\frac{3}{2} & 0 & 0 & 0 \\
0 & \frac{1}{2} & 0 & 0 \\
0 & 0 & -\frac{1}{2} & 0 \\
0 & 0 & 0 & -\frac{3}{2}%
\end{array}%
\right) .
\end{eqnarray}%
These spin operators form an SU(2) Lie algebra $[S^{\alpha },S^{\beta
}]=i\epsilon _{\alpha \beta \gamma }S^{\gamma }$. The first term in Eq. (\ref%
{Hamil}) describes the nearest-neighbor hopping of fermionic atoms, the
second term denotes the on-site Hubbard repulsion between atoms, while the
third term represents the spin-dependent energy of the individual sites.
When the \textit{s}-wave scattering lengths for $S=0$ and $S=2$ are equal,
the third term vanishes and the generalized one-band Hubbard model displays
an SU(4) symmetry \cite{CWu-2003,Hofstetter-2004}. In optical lattices, the
atomic tunneling amplitude $t$ can be easily tuned experimentally by varying
the depth of optical traps. Using the Feshbach resonance, the scattering
lengths $a_{0}$ and $a_{2}$ can be varied, and the coupling parameters $%
c_{0} $ and $c_{2}$ can change over a wide parameter range.

Moreover, for the spin-$\frac{3}{2}$ fermion systems, the spin-quadrupole
operators can be introduced as
\begin{eqnarray}
Q^{xy} &=&\frac{1}{\sqrt{3}}\left( S^{x}S^{y}+S^{y}S^{x}\right) =\Gamma ^{1},
\notag \\
Q^{zx} &=&\frac{1}{\sqrt{3}}\left( S^{z}S^{x}+S^{x}S^{z}\right) =\Gamma ^{2},
\notag \\
Q^{zy} &=&\frac{1}{\sqrt{3}}\left( S^{z}S^{y}+S^{y}S^{z}\right) =\Gamma ^{3},
\notag \\
Q^{(0)} &=&\left( S^{z}\right) ^{2}-\frac{5}{4}=\Gamma ^{4},  \notag \\
Q^{(2)} &=&\frac{1}{\sqrt{3}}\left[ (S^{x})^{2}-(S^{y})^{2}\right] =\Gamma
^{5},
\end{eqnarray}%
where%
\begin{equation*}
\Gamma ^{1}=\left(
\begin{array}{cc}
0 & -iI \\
iI & 0%
\end{array}%
\right) ,\Gamma ^{2,3,4}=\left(
\begin{array}{cc}
\vec{\sigma} & 0 \\
0 & -\vec{\sigma}%
\end{array}%
\right) ,\Gamma ^{5}=\left(
\begin{array}{cc}
0 & I \\
I & 0%
\end{array}%
\right) ,
\end{equation*}%
correspond to the five Dirac matrices, $I$ is a $2\times 2$ unit matrix, and
$\vec{\sigma}_{\alpha }(\alpha =x,y,z)$ are Pauli matrices. The
corresponding dipole and octupole operators can be expressed in terms of
generators of the SO(5) Lie group as $\Gamma ^{ab}=-\frac{i}{2}[\Gamma
^{a},\Gamma ^{b}]$. Thus, three spin operators, five spin-quadrupole
operators, together with seven spin-octupole operators, form the fifteen
generators of the SU(4) Lie group. In terms of the Dirac matrices, the
spin-quadrupole density operator is represented by a five-component vector%
\begin{equation}
n_{i}^{a}=\frac{1}{2}\sum_{\alpha \beta }\psi _{i\alpha }^{\dag }\Gamma
_{\alpha \beta }^{a}\psi _{i\beta }\;(a=1,2,3,4,5).
\end{equation}%
Therefore, the generalized one-band Hubbard model for interacting spin-$%
\frac{3}{2}$ fermions in optical lattices can be rewritten in an SO(5)
invariant form \cite{CWu-2003}%
\begin{eqnarray}
H &=&-t\sum_{\langle ij\rangle ,\alpha }(\psi _{i\alpha }^{\dag }\psi
_{j\alpha }+\text{h.c.})-\frac{3c_{2}}{4}\sum_{i}(\vec{n}_{i})^{2}  \notag \\
&&+\frac{8c_{0}-15c_{2}}{16}\sum_{i}(N_{i}-2)^{2}-\mu \sum_{i}N_{i}.
\end{eqnarray}

At half-filling, the chemical potential $\mu $ should be set to zero to
ensure the particle-hole (p-h) symmetry: $\psi _{i\alpha }\rightarrow
(-1)^{i}\psi _{i\alpha }^{\dag }$, and the average number of fermions per
site should be $\langle N_{i}\rangle =2$. Thus, as far as the Mott
insulating state is concerned, we can safely neglect the particle number
fluctuations focusing on the quantum spin fluctuations. In particular, for $%
c_{2}>0$, \textit{the quantum spin-quadrupole fluctuations are an
interesting feature of the spin-}$\frac{3}{2}$\textit{\ half filled Hubbard
model in the Mott insulating phase.}

\section{Functional integral approach to half filled Hubbard model with $%
c_{2}>0$}

\subsection{Saddle point solution}

It is known that the functional integral approach provides a powerful tool
for studying the antiferromagnetic ground state and spin density wave
excitations of half filled spin-$\frac{1}{2}$ Hubbard model \cite%
{Schulz-1990,Fradkin,ZYWeng-1991,Nagi-1992}. Following the same route, we
will apply the functional integral approach to the generalized spin-$\frac{3%
}{2}$ Hubbard model at half filling.

The partition function is written in an imaginary time functional
path-integral form
\begin{equation}
Z=\int \mathcal{D}\psi ^{\dag }\mathcal{D}\psi \exp \left[ -\int_{0}^{\beta
}L\left( \tau \right) d\tau \right] ,
\end{equation}%
where $\beta =1/T$ and the Lagrangian $L$ is given by $L=\sum_{i,\alpha
}\psi _{i\alpha }^{\dag }\partial _{\tau }\psi _{i\alpha }+H$. In the
following we denote $U\equiv 3c_{2}/4$ for simplicity and then an SO(5)
invariant Hubbard-Stratonovich transformation can be performed,%
\begin{equation*}
Z=\int \mathcal{D}\psi ^{\dag }\mathcal{D}\psi \mathcal{D}\vec{\phi}\exp %
\left[ -\int_{0}^{\beta }d\tau L^{^{\prime }}\left( \tau \right) \right] ,
\end{equation*}%
with%
\begin{eqnarray}
L^{^{\prime }} &=&\sum_{i\alpha }\psi _{i\alpha }^{\dag }\partial _{\tau
}\psi _{i\alpha }-t\sum_{\langle ij\rangle ,\alpha }(\psi _{i\alpha }^{\dag
}\psi _{j\alpha }+\text{H.c.})  \notag \\
&&+\sum_{i}\frac{1}{2}\vec{\phi}_{i}^{2}+\sqrt{\frac{U}{2}}\underset{%
i,\alpha \beta }{\sum }\vec{\phi}_{i}\cdot \psi _{i\alpha }^{\dag }\vec{%
\Gamma}_{\alpha \beta }\psi _{i\beta },
\end{eqnarray}%
where a five-component real bosonic field $\vec{\phi}_{i}(\tau )$ has been
introduced. By integrating out the fermion fields $\psi ^{\dag }$ and $\psi $%
, we obtain an effective action%
\begin{equation}
S_{\text{eff}}=\int_{0}^{\beta }d\tau \sum_{i}\frac{1}{2}\vec{\phi}%
_{i}^{2}\left( \tau \right) -\text{Tr}\ln \left[ \partial _{\tau }+\mathbf{M}%
\right] .
\end{equation}%
Here the trace is taken over the Nambu space, the spatial and imaginary time
coordinates. The matrix element of $\mathbf{M}$ and fermionic Green's
function (GF) are given by
\begin{eqnarray}
G_{\alpha \beta }(\mathbf{r}_{i},\tau ;\mathbf{r}_{j},\tau ^{\prime })
&=&-\langle \mathbf{r}_{i},\tau ,\alpha |\frac{1}{\partial _{\tau }+\mathbf{M%
}}|\mathbf{r}_{j},\tau ^{\prime },\beta \rangle ,  \notag \\
\langle \mathbf{r}_{i},\tau ,\alpha |\mathbf{M}|\mathbf{r}_{j},\tau ^{\prime
},\beta \rangle &=&-2t\delta _{\alpha \beta }\delta _{\tau \tau ^{\prime
}}\delta _{i,j+\delta }  \notag \\
&&+\sqrt{\frac{U}{2}}\delta _{\tau \tau ^{\prime }}\delta _{ij}\vec{\phi}%
_{i}(\tau )\cdot \vec{\Gamma}_{\alpha \beta }.
\end{eqnarray}%
So far no approximations have been made.

In order to reveal the consequences of the spin-quadrupole fluctuations, we
first consider the saddle-point solution of the effective action.
Differentiating $S_{\text{eff}}$ with respect to $\phi _{i}^{a}\left( \tau
\right) $, we obtain%
\begin{equation}
\phi _{i}^{a}\left( \tau \right) =-\sqrt{\frac{U}{2}}\underset{\alpha \beta }%
{\sum }G_{\alpha \beta }(\mathbf{r}_{i},\tau ;\mathbf{r}_{i},\tau )\Gamma
_{\beta \alpha }^{a},
\end{equation}%
corresponding to a mean field result \cite{CWu-2003}. It is expected that
the lowest energy state is given by a staggered phase of the SO(5) vector,
namely, a staggered spin-quadrupole (spin nematic) ordered phase with an
order parameter $\vec{\phi}_{i}\left( \tau \right) \rightarrow |\vec{\phi}%
|e^{i\mathbf{Q}\cdot \mathbf{r}_{i}}\hat{d}$ , where $\hat{d}$
vector corresponds to one of the five spin-quadrupole components
resulting from spontaneous symmetry breaking of SO(5) to SO(4). On
a two-dimensional (2D) square lattice $\mathbf{Q}=(\pi ,\pi )$
corresponds to the reciprocal wave vector. Compared with the
conventional magnetic long-range ordered states, there is no time
reversal symmetry breaking in the spin-quadrupole ordering state.
Then the effective action at the saddle point becomes
\begin{widetext}
\begin{equation}
S^{\prime }=\sum_{\mathbf{k},i\mathbf{\omega }_{n},\alpha \beta }\left( \psi _{\alpha }^{\dag }(\mathbf{k},i%
\mathbf{\omega }_{n}),\psi _{\alpha }^{\dag }(\mathbf{k-Q},i\mathbf{\omega }%
_{n})\right) \left(
\begin{array}{cc}
\left( -i\mathbf{\omega }_{n}+\varepsilon _{\mathbf{k}}\right)
\delta _{\alpha \beta } &
\sqrt{\frac{U}{2}}|\vec{\phi}|\vec{\Gamma}_{\alpha \beta
}\cdot \hat{d} \\
\sqrt{\frac{U}{2}}|\vec{\phi}|\vec{\Gamma}_{\alpha \beta }\cdot
\hat{d} & \left( -i\mathbf{\omega }_{n}-\varepsilon
_{\mathbf{k}}\right) \delta
_{\alpha \beta }%
\end{array}%
\right) \left(
\begin{array}{c}
\psi _{\beta }(\mathbf{k},i\mathbf{\omega }_{n}) \\
\psi _{\beta }(\mathbf{k-Q},i\mathbf{\omega }_{n})%
\end{array}%
\right) +\frac{1}{2}\beta N|\vec{\phi}|^{2},
\end{equation}%
\end{widetext}
where the summation over momentum is limited to the reduced Brillouin zone,
the dispersion relation is $\varepsilon _{\mathbf{k}}=-2t(\cos k_{x}+$ $\cos
k_{y})$, and $\mathbf{\omega }_{n}$ is the fermionic Matsubara frequency.

Due to the doubled unit cell, the fermion GF generally has to be defined as $%
G_{\alpha \beta }\left( \mathbf{k},\mathbf{k}^{\prime };\tau \right)
=-\langle T_{\tau }\psi _{\mathbf{k}\alpha }(\tau )\psi _{-\mathbf{k}%
^{\prime }\beta }^{\dag }(0)\rangle $, which has non-zero off-diagonal terms
in momentum space due to the umklapp processes with respect to $\mathbf{Q}$.
Explicitly, the expression of the single-particle GF can be written as%
\begin{eqnarray}
&&G_{\alpha \beta }\left( \mathbf{k},-\mathbf{k}^{\prime };i\omega
_{n}\right)  \notag \\
&=&\frac{\left( i\omega _{n}+\varepsilon _{\mathbf{k}}\right) \delta
_{\alpha \beta }\delta _{\mathbf{kk}^{\prime }}+\sqrt{\frac{U}{2}}|\vec{\phi}%
|(\vec{\Gamma}_{\alpha \beta }\cdot \hat{d})\delta _{\mathbf{k}^{\prime },%
\mathbf{k}-\mathbf{Q}}}{(i\omega _{n})^{2}-E_{\mathbf{k}}^{2}},
\end{eqnarray}%
where GF poles lead to the quasiparticle spectra $E_{\mathbf{k}}=\pm \sqrt{%
\varepsilon _{\mathbf{k}}^{2}+\frac{U}{2}|\vec{\phi}|^{2}}$. At half
filling, the upper band is empty while the lower band\ is completely filled.
Thus an energy gap $\Delta =2\sqrt{\frac{U}{2}}|\vec{\phi}|$ opens up in the
quasiparticle spectrum. By Fourier transformation, the gap equation is given
by%
\begin{equation}
1-\frac{2U}{\beta N}\underset{\mathbf{k},i\omega _{n}}{\sum }\frac{1}{\omega
_{n}^{2}+E_{\mathbf{k}}^{2}}=0.
\end{equation}%
At $T=0$K, for arbitrarily small $U$ there is always a finite energy gap.
Particularly, in the limit of $U\ll t$, it gives rise to
\begin{equation}
\Delta \simeq 2te^{-\pi \sqrt{2t/U}}.
\end{equation}%
However, for $U\gg t$, we have $\Delta \simeq 2U$, i.e., the Mott gap in the
single-particle excitations.

Moreover, the sublattice spin-quadrupole moment is related to the energy gap
and is given by%
\begin{equation}
|\vec{n}_{i}|=\frac{\Delta }{2U}.
\end{equation}%
We have numerically calculated the energy gap and plotted the
spin-quadrupole moment in Fig.1. For a large value of $U$, $\Delta \simeq 2U$
and $|\vec{n}_{i}|\rightarrow 1$, i.e., a saturated spin-quadrupole moment.
\begin{figure}[tbp]
\includegraphics[scale=0.7]{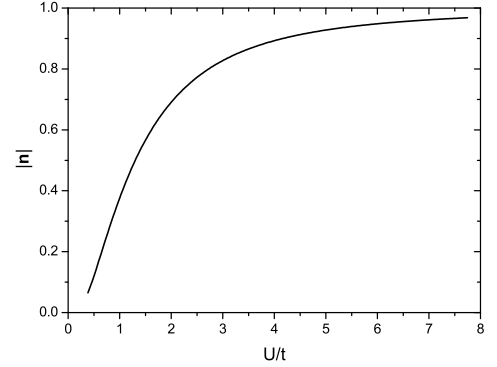}
\caption{The spin-quadrupole moment $|\vec{n}_{i}|=\frac{\Delta }{2U}$
obtained in the saddle point approximation.}
\end{figure}

\subsection{Spin-quadrupole correlations at the saddle point}

The longitudinal and transverse spin-quadrupole correlation functions are
defined as
\begin{eqnarray}
\chi _{\sigma }(\mathbf{p},\mathbf{p}^{\prime };\tau ) &=&\frac{1}{N}\langle
\text{T}_{\tau }[\vec{n}_{\mathbf{p}}(\tau )\cdot \hat{d}][\vec{n}_{-\mathbf{%
p}^{\prime }}(0)\cdot \hat{d}]\rangle ,  \notag \\
\chi _{\pi \pi ^{\prime }}(\mathbf{p},\mathbf{p}^{\prime };\tau ) &=&\frac{1%
}{N}\langle \text{T}_{\tau }[\vec{n}_{\mathbf{p}}(\tau )\cdot \hat{e}_{n}][%
\vec{n}_{-\mathbf{p}^{\prime }}(0)\cdot \hat{e}_{n^{\prime }}]\rangle ,
\end{eqnarray}%
where $\hat{d}$ corresponds to the spin-quadrupole ordering direction, and $%
\hat{e}_{n}$ and $\hat{e}_{n^{\prime }}$ are unit vectors perpendicular to $%
\hat{d}$ vector. At the saddle point, the direction of $\hat{d}$ and $\hat{e}%
_{n}$ does not change in time and space, and these correlation functions are
expressed in terms of the single-particle GFs by using a Nambu spinor $\psi
_{i}=\left( \psi _{i,3/2},\psi _{i,1/2},\psi _{i,-1/2},\psi _{i,-3/2}\right)
^{T}$. Actually, we notice that the transverse correlation $\chi _{\pi \pi
^{\prime }}(\mathbf{p},\mathbf{p}^{\prime };\tau )$ vanishes when $\hat{e}%
_{n}\perp \hat{e}_{n^{\prime }}$, and yields a finite value only when $\hat{e%
}_{n}=\hat{e}_{n^{\prime }}$. Hence the correlation functions are written as
\begin{eqnarray*}
\chi _{\sigma }(\mathbf{p},\mathbf{p}^{\prime };i\omega _{l}) &=&-\frac{1}{%
4\beta N}\underset{\mathbf{kk}^{\prime },i\omega _{n}}{\sum }\text{Tr}\left[
(\mathbf{\Gamma \cdot \hat{d}})\mathbf{G}(\mathbf{k},-\mathbf{k}^{\prime
};i\omega _{n})\right. \\
&&\times \left. (\mathbf{\Gamma \cdot \hat{d}})\mathbf{G}(\mathbf{k}^{\prime
}+\mathbf{p}^{\prime },-\mathbf{k}-\mathbf{p};i\omega _{n}+i\omega _{l})%
\right] , \\
\chi _{\pi }(\mathbf{p},\mathbf{p}^{\prime };i\omega _{l}) &=&-\frac{1}{%
4\beta N}\underset{\mathbf{kk}^{\prime },i\omega _{n}}{\sum }\text{Tr}\left[
(\mathbf{\Gamma \cdot \hat{e}}_{n})G(\mathbf{k},-\mathbf{k}^{\prime
};i\omega _{n})\right. \\
&&\times \left. (\mathbf{\Gamma \cdot \hat{e}}_{n})\mathbf{G}(\mathbf{k}%
^{\prime }+\mathbf{p}^{\prime },-\mathbf{k}-\mathbf{p};i\omega _{n}+i\omega
_{l})\right] .
\end{eqnarray*}%
After some straightforward algebra, the spin-quadrupole correlation
functions are expressed as%
\begin{eqnarray}
&&\binom{\chi _{\sigma }(\mathbf{p},\mathbf{p}^{\prime };i\omega _{l})}{\chi
_{\pi }(\mathbf{p},\mathbf{p}^{\prime };i\omega _{l})}  \notag \\
&=&\frac{\delta _{\mathbf{pp}^{\prime }}}{\beta N}\underset{\mathbf{k}%
,i\omega _{n}}{\sum }\frac{\omega _{n}\left( \omega _{n}+\omega _{l}\right)
-\varepsilon _{\mathbf{k}}\varepsilon _{\mathbf{k+p}}\mp \frac{\Delta ^{2}}{4%
}}{(\omega _{n}^{2}+E_{\mathbf{k}}^{2})\left[ \left( \omega _{n}+\omega
_{l}\right) ^{2}+E_{\mathbf{k}+\mathbf{p}}^{2}\right] }.  \label{quadrupole}
\end{eqnarray}%
Performing the Matsubara frequency summation and analytical continuation,
the imaginary part of $\chi (\mathbf{p},\mathbf{p}^{\prime };\omega )$ can
be derived. Since the off-diagonal terms in momentum space vanish, we can
only consider $\chi (\mathbf{p},\mathbf{p};\omega )$. To display the spatial
correlations, the static spin-quadrupole structure factors are evaluated
through the fluctuation-dissipation relation $S(\mathbf{p})=\frac{1}{\pi }%
\int_{-\infty }^{+\infty }d\omega \lbrack 1+n_{B}(\omega )]$Im$\chi (\mathbf{%
p},\mathbf{p};\omega )$, where $n_{B}(\omega )$ is the Bose-Einstein
distribution function. At $T=0$K, the static structure factors are obtained
\begin{equation}
\binom{S_{\sigma }(\mathbf{p})}{S_{\pi }(\mathbf{p})}=\frac{1}{4N}\underset{%
\mathbf{k}}{\sum }(1-\frac{\varepsilon _{\mathbf{k}}\varepsilon _{\mathbf{k}+%
\mathbf{p}}\pm \frac{\Delta ^{2}}{4}}{E_{\mathbf{k}}E_{\mathbf{k}+\mathbf{p}}%
}).
\end{equation}%
The static spin-quadrupole structure factors are numerically calculated and
displayed in Fig.2, where broad peaks appear at $\mathbf{p}=(\pi ,\pi )$,
indicating antiferroquadrupolar long-range correlations.
\begin{figure}[tbp]
\includegraphics[scale=0.5]{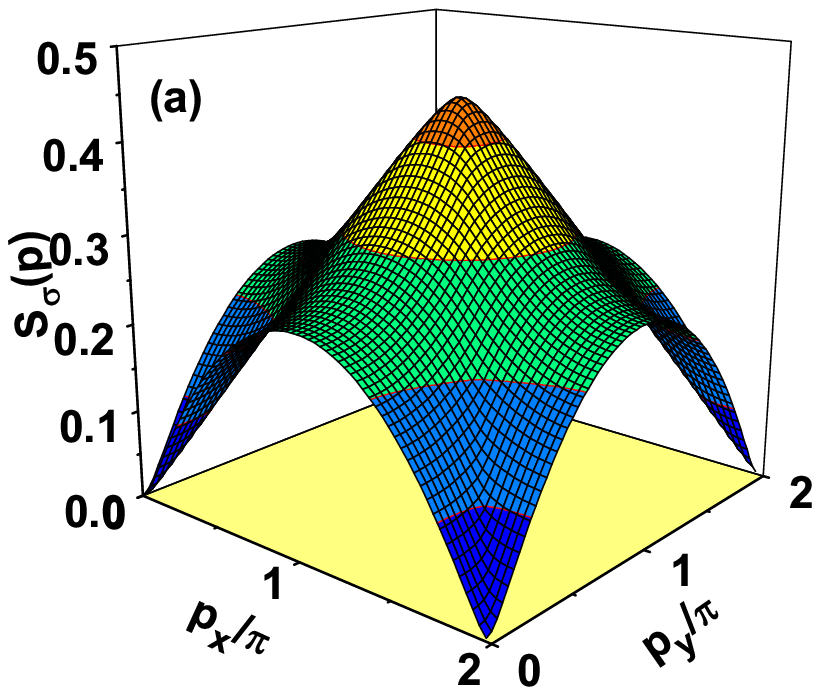} %
\includegraphics[scale=0.5]{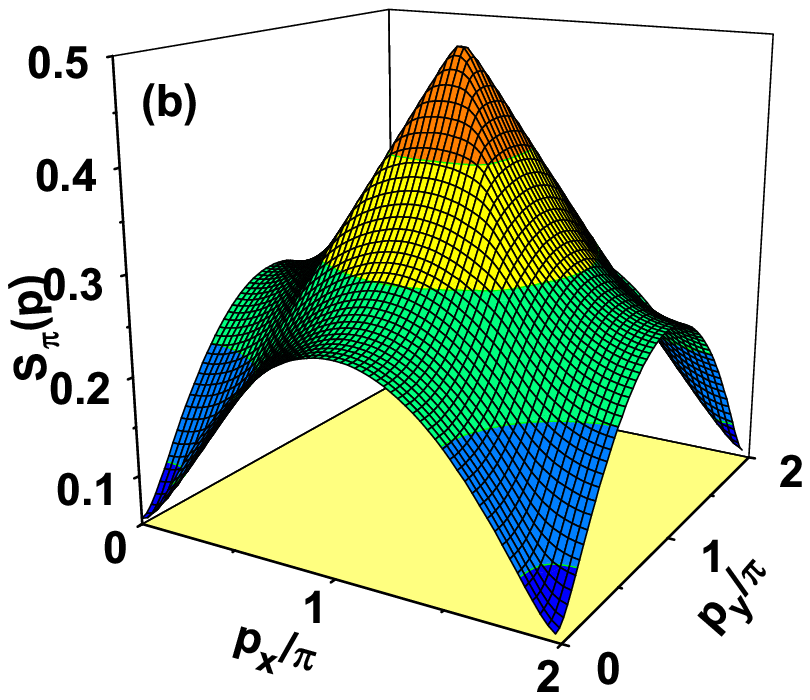}
\caption{(Color online) Longitudinal static spin-quadrupole structure factor
$S_{\protect\sigma }(\mathbf{p})$ (a) and transverse static spin-quadrupole
structure factor $S_{\protect\pi }(\mathbf{p})$ (b) in the 2D staggered spin
nematic ordered phase with $U/t=0.686$ and $\Delta =0.3$.}
\end{figure}

\subsection{Anisotropic spin fluctuations at the saddle point}

It is interesting to examine the spin-spin correlation functions in the
presence of staggered spin-quadrupolar ordering background. Let us first
look at the dipole moment,%
\begin{equation}
\langle S_{i}^{\gamma }\rangle =-\frac{1}{\beta N}\underset{\mathbf{k}%
,i\omega _{n}}{\sum }\text{Tr}\left[ S^{\gamma }\mathbf{G}(\mathbf{k},%
\mathbf{Q}-\mathbf{k};i\omega _{n})\right] =0,
\end{equation}%
where $S^{\gamma }$ $(\gamma =x,y,z)$ denotes the corresponding spin-$\frac{3%
}{2}$ matrices. This result implies that\textit{\ there is no
spin-dipole long-range order in the saddle point solution}. To get
the collective
excitations, the spin-correlation functions have to be evaluated,%
\begin{equation}
\chi ^{\alpha \beta }(\mathbf{p},\mathbf{p}^{\prime };\tau )=\frac{1}{N}%
\langle \text{T}_{\tau }S_{\mathbf{p}}^{\alpha }(\tau )S_{-\mathbf{p}%
^{\prime }}^{\beta }(0)\rangle .
\end{equation}%
By expressing the spin-density operators in terms of Nambu spinor as $S_{%
\mathbf{p}}^{\gamma }=\sum_{\mathbf{k},\alpha \beta }\psi _{\mathbf{k}+%
\mathbf{p},\alpha }^{\dag }S_{\alpha \beta }^{\gamma }\psi _{\mathbf{k}\beta
}$, the spin-correlation functions are given by%
\begin{eqnarray}
\chi ^{\alpha \beta }(\mathbf{p},\mathbf{p}^{\prime };i\omega _{l}) &=&-%
\frac{1}{\beta N}\underset{\mathbf{kk}^{\prime },i\omega _{n}}{\sum }\text{Tr%
}\left[ S^{\alpha }\mathbf{G}(\mathbf{k},-\mathbf{k}^{\prime };i\omega
_{n})\right.  \notag \\
&&\times \left. S^{\beta }\mathbf{G}(\mathbf{k}^{\prime }+\mathbf{p}^{\prime
},-\mathbf{k}-\mathbf{p};i\omega _{n}+i\omega _{l})\right] .
\end{eqnarray}%
\begin{figure}[tbp]
\centering \includegraphics [width=6cm]{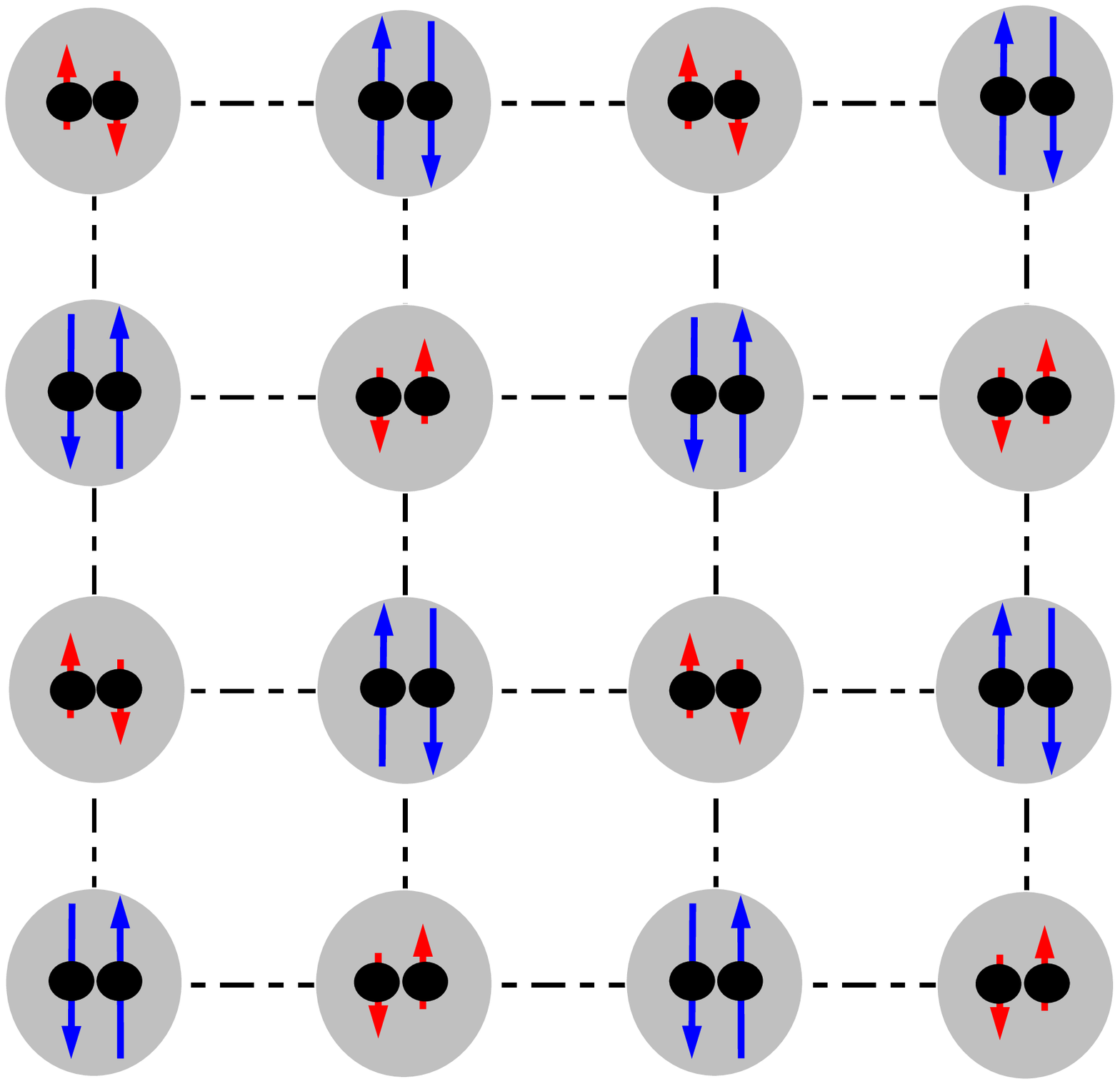} \caption{The
picture of the spin-quadrupole long-range ordered state with the
order parameter $(-1)^{i}n_{i}^{4}$ on a two-dimensional square
lattice, where each lattice site is occupied by a spin singlet
state.}
\end{figure}

Without loss of generality, we may assume the spin-quadrupole ordering is
along $\hat{d}=\hat{e}_{4}$. Then the order parameter is denoted by $%
(-1)^{i}\langle n_{i}^{4}\rangle \neq 0$, where
\begin{equation}
2n_{i}^{4}=\psi _{i,\frac{3}{2}}^{\dagger }\psi _{i,\frac{3}{2}}+\psi _{i,-%
\frac{3}{2}}^{\dagger }\psi _{i,-\frac{3}{2}}-\psi _{i,\frac{1}{2}}^{\dagger
}\psi _{i,\frac{1}{2}}-\psi _{i,-\frac{1}{2}}^{\dagger }\psi _{i,-\frac{1}{2}%
},
\end{equation}%
corresponding to the difference between the $S^{z}=\pm 3/2$ and $S^{z}=\pm
1/2$ spin densities in the site singlet state. The physical picture
corresponding to this particular spin-quadrupole ordering state is displayed
in Fig.3. Inserting the fermionic GF into the above expression of the spin
correlations, we find that%
\begin{equation}
\chi ^{xx}(\mathbf{p},\mathbf{p}^{\prime };i\omega _{l})=\chi ^{yy}(\mathbf{p%
},\mathbf{p}^{\prime };i\omega _{l})\neq \chi ^{zz}(\mathbf{p},\mathbf{p}%
^{\prime };i\omega _{l}),
\end{equation}%
which implies that in the spin-quadrupole long-range ordered state \textit{%
the time reversal symmetry is reserved but the spin rotational symmetry is
broken. }This is the main characteristics of the spin-quadrupole or spin
nematic ordered state. Thus the corresponding expressions of the
spin-correlation functions are written as%
\begin{eqnarray}
&&\chi ^{zz}(\mathbf{p},\mathbf{p}^{\prime };i\omega _{l})  \notag \\
&=&\frac{5\delta _{\mathbf{pp}^{\prime }}}{\beta N}\underset{\mathbf{k}%
,i\omega _{n}}{\sum }\frac{\omega _{n}\left( \omega _{n}+\omega _{l}\right)
-\varepsilon _{\mathbf{k}}\varepsilon _{\mathbf{k+p}}-\frac{\Delta ^{2}}{4}}{%
(\omega _{n}^{2}+E_{\mathbf{k}}^{2})\left[ \left( \omega _{n}+\omega
_{l}\right) ^{2}+E_{\mathbf{k}+\mathbf{p}}^{2}\right] }, \\
&&\chi ^{+-}(\mathbf{p},\mathbf{p}^{\prime };i\omega _{l})  \notag \\
&=&\frac{2\delta _{\mathbf{pp}^{\prime }}}{\beta N}\underset{\mathbf{k}%
,i\omega _{n}}{\sum }\frac{5\omega _{n}\left( \omega _{n}+\omega _{l}\right)
-5\varepsilon _{\mathbf{k}}\varepsilon _{\mathbf{k+p}}+\frac{\Delta ^{2}}{4}%
}{(\omega _{n}^{2}+E_{\mathbf{k}}^{2})\left[ \left( \omega _{n}+\omega
_{l}\right) ^{2}+E_{\mathbf{k}+\mathbf{p}}^{2}\right] }.
\end{eqnarray}%
It should be emphasized that unlike the half filled spin-$\frac{1}{2}$
Hubbard model \cite{SDW}, all off-diagonal spin correlations in the momentum
space vanish, which will be discussed in the next subsection. At $T=0K$, we
may rewrite $\chi (\mathbf{p},\mathbf{p}^{\prime };\omega )=\chi (\mathbf{p}%
,\omega )\delta _{\mathbf{pp}^{\prime }}$, and the static spin structure
factors can be derived as
\begin{eqnarray}
S^{zz}(\mathbf{p}) &=&\frac{5}{4N}\underset{\mathbf{k}}{\sum }(1-\frac{%
\varepsilon _{\mathbf{k}}\varepsilon _{\mathbf{k}+\mathbf{p}}+\Delta ^{2}/4}{%
E_{\mathbf{k}}E_{\mathbf{k}+\mathbf{p}}}),  \notag \\
S^{+-}(\mathbf{p}) &=&\frac{1}{2N}\underset{\mathbf{k}}{\sum }(9-\frac{%
5\varepsilon _{\mathbf{k}}\varepsilon _{\mathbf{k}+\mathbf{p}}-\Delta ^{2}/4%
}{E_{\mathbf{k}}E_{\mathbf{k}+\mathbf{p}}}).
\end{eqnarray}%
\begin{figure}[tbp]
\includegraphics[scale=0.5]{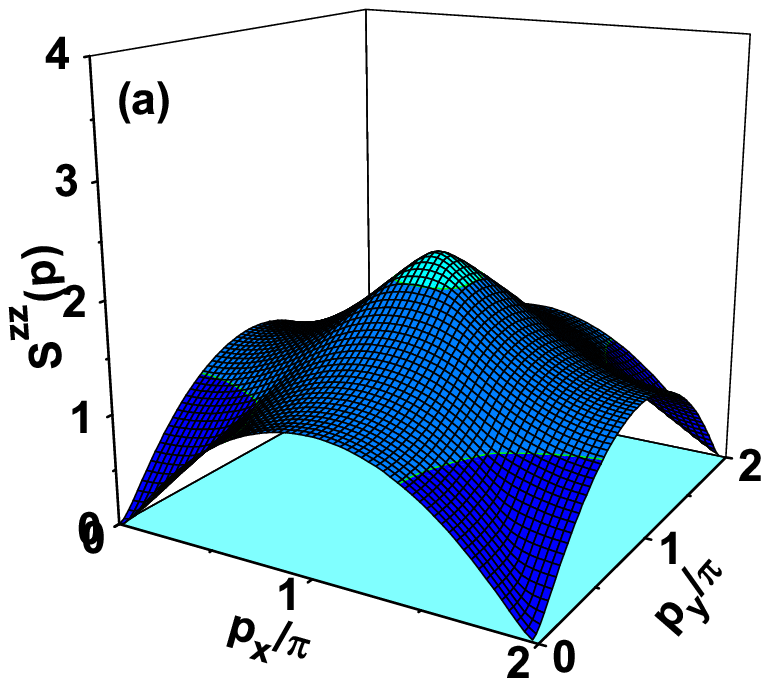} \includegraphics[scale=0.5]{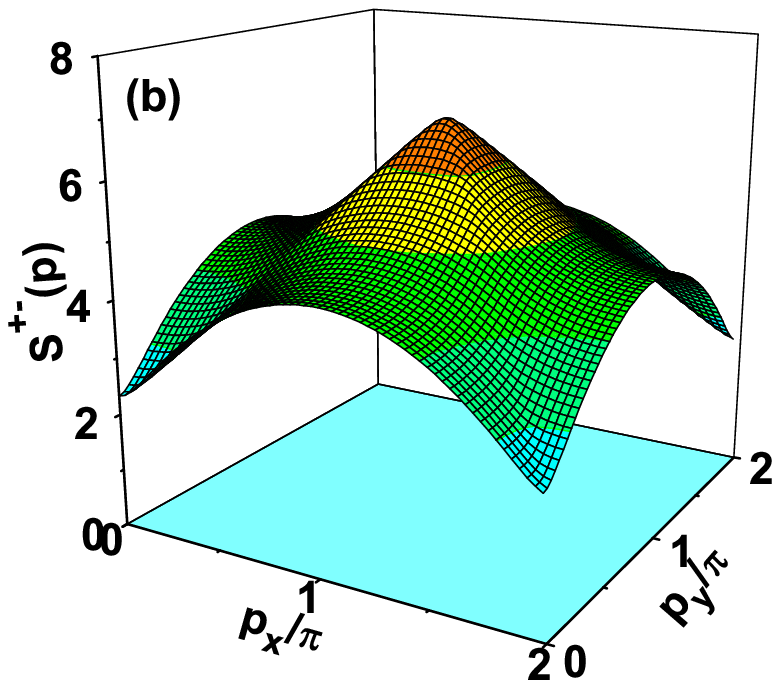}
\caption{(Color online) Static spin structure factor $S^{zz}(\mathbf{p})$
(a) and $S^{+-}(\mathbf{p})$ (b) in 2D staggered spin nematic ordered phase
with $U/t=0.686$ and $\Delta =0.3$.}
\end{figure}

We have plotted the corresponding numerical results in Fig.4. For both $%
S^{zz}(\mathbf{p})$ \ and $S^{+-}(\mathbf{p})$, much broader peaks at the
momentum $(\pi ,\pi )$ indicate the presence of strong antiferromagnetic
fluctuations. The weaker broad peaks at $(0,\pi )$ and $(\pi ,0)$ show
collinear spin-dimer correlations. Actually, these peculiar features of spin
and spin-quadrupole structure factors can be detected by polarized light
Bragg scattering or spatial quantum noise interferometry experiments \cite%
{Bloch-2004}, which are reliable experimental probes to detect magnetic
correlations of ultracold atoms in optical lattices.

\subsection{Spin-quadrupole density waves}

To further study the spin-quadrupole collective excitation modes, we
consider Gaussian fluctuations around the saddle point solution, $\vec{\phi}=%
\vec{\phi}_{c}+\delta \vec{\phi}$. Then, according to%
\begin{eqnarray}
\text{Tr}\ln \left[ \partial _{\tau }+\mathbf{M}\right] &=&\text{Tr}\ln %
\left[ -\mathbf{G}^{-1}(1-\mathbf{GV})\right]  \notag \\
&=&\text{Tr}\ln (-\mathbf{G}^{-1})-\sum_{n}\frac{1}{n}\text{Tr}(\mathbf{GV}%
)^{n},
\end{eqnarray}%
the effective action $S_{\text{eff}}$ can be expanded as $S_{\text{eff}%
}=\sum_{n=0}^{\infty }S^{(n)}(\vec{\phi}_{c},\delta \vec{\phi})$, where $%
\mathbf{G}$ represents the fermionic GF at the saddle point and the matrix
element of $\mathbf{V}$ is given by%
\begin{equation}
\langle \mathbf{r}_{i},\tau ,\alpha |\mathbf{V}|\mathbf{r}_{j},\tau ^{\prime
},\beta \rangle =\delta _{\tau \tau ^{\prime }}\delta _{ij}\sqrt{\frac{U}{2}}%
\delta \vec{\phi}_{i}(\tau )\cdot \vec{\Gamma}_{\alpha \beta }.
\end{equation}%
Since $\mathbf{V}$ only contains a linear term in $\delta
\vec{\phi}$, the above procedure is indeed an expansion in the
spin-quadrupole fluctuation field $\delta \vec{\phi}$. The
first-order term in $\delta \vec{\phi}$ vanishes due to the saddle
point condition. After some algebra, up to the second-order
expansion, we arrive at
\begin{eqnarray}
S^{(2)}(\delta \vec{\phi}) &=&\frac{1}{2}\sum_{\mathbf{pp}^{\prime },i\omega
_{l}}\sum_{ab}\delta \phi ^{a}\left( \mathbf{p}^{\prime },-i\omega
_{l}\right) \delta \phi ^{b}\left( \mathbf{p},i\omega _{l}\right)  \notag \\
&&\times K^{ab}\left( \mathbf{p},\mathbf{p}^{\prime };i\omega _{l}\right) ,
\end{eqnarray}%
where%
\begin{eqnarray}
&&K^{ab}(\mathbf{p},\mathbf{p}^{\prime };i\omega _{l})  \notag \\
&=&\delta _{ab}\delta _{\mathbf{p}^{\prime },-\mathbf{p}}+\frac{U}{2\beta N}%
\underset{\mathbf{kk}^{\prime },i\omega _{n}}{\sum }\text{Tr}\left[ \mathbf{G%
}(\mathbf{k},-\mathbf{k}^{\prime };i\omega _{n})\right.  \notag \\
&&\text{ \ \ \ }\times \left. \Gamma ^{a}\mathbf{G}(\mathbf{k}^{\prime }-%
\mathbf{p}^{\prime },-\mathbf{k}-\mathbf{p};i\omega _{n}+i\omega _{l})\Gamma
^{b}\right] ,
\end{eqnarray}%
$1\leq a,b\leq 5$ and $\omega _{l}$ are bosonic Matsubara frequencies. With
the help of the following relations for the Dirac gamma matrices%
\begin{eqnarray}
\text{Tr}1 &=&4,\ \ \text{Tr}\Gamma ^{a}=0,  \notag \\
\text{Tr}\left( \Gamma ^{a}\Gamma ^{b}\right) &=&4\delta _{ab},\qquad \text{%
Tr}\left( \Gamma ^{a}\Gamma ^{b}\Gamma ^{c}\right) =0,  \notag \\
\text{Tr}\left( \Gamma ^{a}\Gamma ^{b}\Gamma ^{c}\Gamma ^{d}\right)
&=&4\left( \delta _{ab}\delta _{cd}-\delta _{ac}\delta _{bd}+\delta
_{ad}\delta _{bc}\right) ,
\end{eqnarray}%
the above kernel functions are expressed as%
\begin{equation*}
K^{ab}(\mathbf{p},\mathbf{p}^{\prime },i\omega _{l})=[K_{0}\left( \mathbf{p}%
,i\omega _{l}\right) \delta _{ab}+K_{1}\left( \mathbf{p},i\omega _{l}\right)
d_{a}d_{b}]\delta _{\mathbf{p}^{\prime },-\mathbf{p}},
\end{equation*}%
with
\begin{eqnarray}
K_{0}(\mathbf{p},i\omega _{l}) &=&1-2U\chi _{\pi }(\mathbf{p},\mathbf{p}%
;i\omega _{l}),  \notag \\
K_{1}(\mathbf{p},i\omega _{l}) &=&2U[\chi _{\sigma }(\mathbf{p},\mathbf{p}%
;i\omega _{l})-\chi _{\pi }(\mathbf{p},\mathbf{p};i\omega _{l})],
\end{eqnarray}%
where $\chi _{\sigma }(\mathbf{p},\mathbf{p};i\omega _{l})$ and $\chi _{\pi
}(\mathbf{p},\mathbf{p};i\omega _{l})$ correspond to the spin-quadrupole
correlation functions Eq.(\ref{quadrupole}) derived at the saddle-point
approximation.

Decomposing the spin-quadrupole fluctuation field $\delta \vec{\phi}$ into
one longitudinal component $\sigma $ parallel to the direction $\hat{d}$ and
four transverse components $\vec{\pi}$ perpendicular to the direction $\hat{d%
}$, i.e., $\delta \vec{\phi}=\binom{\sigma \hat{d}}{\vec{\pi}}$. In terms of
$\sigma $ and $\vec{\pi}$, the effective action with Gaussian fluctuations
is expressed as%
\begin{eqnarray}
S^{(2)}(\delta \vec{\phi}) &=&\sum_{\mathbf{p},i\omega _{l}}\frac{1}{2}\left[
K_{0}(\mathbf{p},i\omega _{l})+K_{1}(\mathbf{p},i\omega _{l})\right] \left|
\sigma \left( \mathbf{p},i\omega _{l}\right) \right| ^{2}  \notag \\
&&+\sum_{\mathbf{p},i\omega _{l}}\frac{1}{2}K_{0}(\mathbf{p},i\omega _{l})|%
\vec{\pi}(\mathbf{p},i\omega _{l})|^{2}.
\end{eqnarray}%
In the low energy limit $i\omega _{l}\rightarrow 0$, for the momentum
transfer $\mathbf{p}=\mathbf{Q}$, we find $K_{0}\left( \mathbf{Q},0\right)
=0 $ and $K_{1}\left( \mathbf{Q},0\right) >0$ following from the
quasiparticle gap equation. Hence the corresponding four transverse modes $%
\vec{\pi}$ are nothing but the Goldstone boson modes induced by the
spontaneous symmetry breaking from SO(5) to SO(4), living on the space $%
SO(5)/SO(4)=S^{4}$, while the $\sigma $ mode is gapful. Thus we identify
these Goldstone collective excitation modes with spin-quadrupole density
waves.

For the spin-density wave (SDW) in the half filled
spin-$\frac{1}{2}$
Hubbard model, it had been noticed that the coupled vibrations of $S_{x}(%
\mathbf{q})$ and $S_{y}(\mathbf{Q}+\mathbf{q})$ produce the spin density
wave \cite{SDW,Halperin-1969}. In the functional integral approach, up to
the Gaussian fluctuations, the coupled transverse vibrations are generated
and allowed by the symmetry breaking of SO(3) to SO(2) in the effective
action \cite{Fradkin,Nagi-1992}, which is also the manifestation of an
identity for Pauli matrices: Tr$\left( \sigma ^{a}\sigma ^{b}\sigma
^{c}\right) =2i\epsilon ^{abc}$. Due to such coupled transverse vibration
modes, the SDW velocity is strongly suppressed in the limit of $U\gg t$, and
there exist the off-diagonal transverse spin-correlation functions in the
momentum space. To the contrary, we have found that the transverse mode
couplings in spin-quadrupole density waves are absent, because the SO(4)
invariance of the effective action under the Gaussian fluctuations does not
allow such couplings. This is also the consequence of an identity for the
Dirac gamma matrices: Tr$\left( \Gamma ^{a}\Gamma ^{b}\Gamma ^{c}\right) =0$%
. Actually, the same reason leads to vanishing of the off-diagonal terms of
the transverse spin-quadrupole density correlations in the momentum space,
as pointed out in the last subsection. The absence of the transverse
spin-quadrupole density wave mode coupling is a remarkable property of the
enlarged hyperfine spin space dimensionality and the higher symmetry of the
local interactions in the generalized Hubbard model.

In order to evaluate the spin-quadrupole density wave velocity, we perform
the Matsubara frequency summation of $K_{0}\left( \mathbf{Q}+\mathbf{q}%
,\omega \right) $ and analytical continuation, where $\mathbf{q}$ is small
and $\omega \rightarrow 0$. At $T=0$, $K_{0}\left( \mathbf{Q}+\mathbf{q}%
,\omega \right) $ is expanded up to the second order in $\mathbf{q}$ and $%
\omega $, namely,
\begin{equation}
K_{0}(\mathbf{Q}+\mathbf{q},\omega )\approx a\mathbf{q}^{2}-b\omega ^{2},
\end{equation}%
with%
\begin{equation}
a=\frac{Ut^{2}}{N}\underset{\mathbf{k}}{\sum }\frac{\sin ^{2}k_{x}}{E_{%
\mathbf{k}}^{3}},\qquad b=\frac{U}{4N}\underset{\mathbf{k}}{\sum }\frac{1}{%
E_{\mathbf{k}}^{3}}.
\end{equation}%
Thus, the effective action of $\vec{\pi}$ mode can be expressed as%
\begin{equation}
S^{(2)}(\vec{\pi})=\frac{1}{2}\int \frac{d\mathbf{q}d\omega }{(2\pi )^{3}}%
\rho (v_{s}^{2}\mathbf{q}^{2}-\omega ^{2})\left\vert \vec{\pi}\left( \mathbf{%
q},\omega \right) \right\vert ^{2},
\end{equation}%
with spin-quadrupole density wave stiffness $\rho =b$ and velocity $v_{s}=%
\sqrt{a/b}$. The corresponding numerical results are shown in Fig.5. In the
small $U$ limit, the spin-quadrupole wave velocity is approximated by%
\begin{equation}
v_{s}\approx \frac{2t}{\sqrt{\pi }}(\frac{2U}{t})^{1/4},
\end{equation}%
while in the large $U$ limit it is given by
\begin{equation}
v_{s}\approx \sqrt{2}t(1+\frac{3t^{2}}{4U^{2}}).
\end{equation}%
Due to the absence of the coupled vibrations of the transverse modes, the
spin-quadrupole density wave velocity is \textit{saturated in the strong
coupling limit}, in a \textit{sharp contrast to the spin-}$\frac{1}{2}$%
\textit{\ case}. These properties are based on the symmetry consideration.
Therefore, the above results obtained in the Gaussian approximation around
the saddle point should be valid even in the presence of higher-order
fluctuations.
\begin{figure}[tbp]
\includegraphics[scale=0.45]{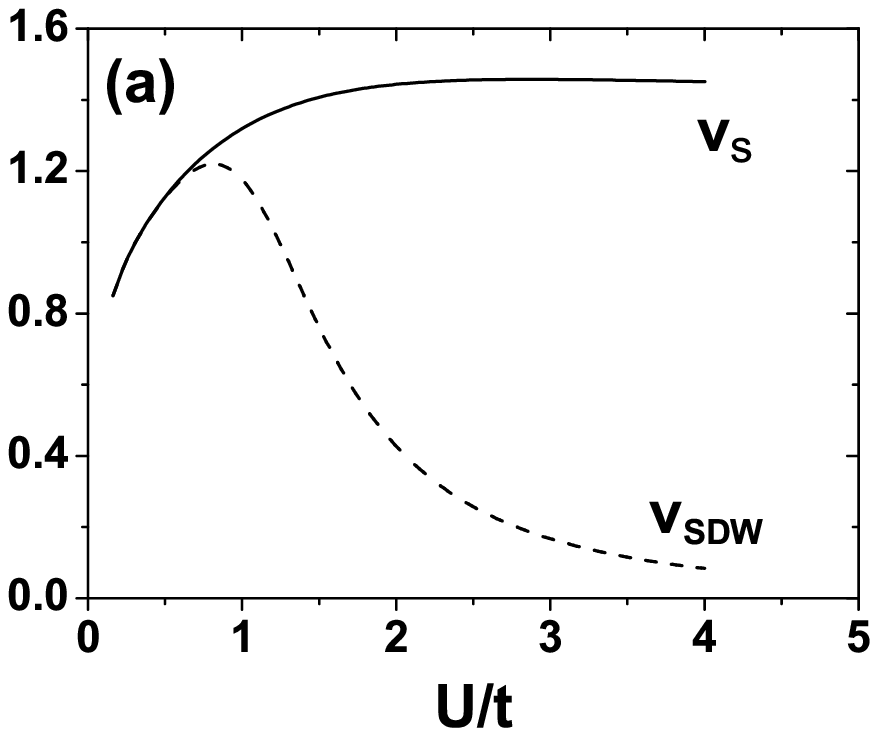} %
\includegraphics[scale=0.45]{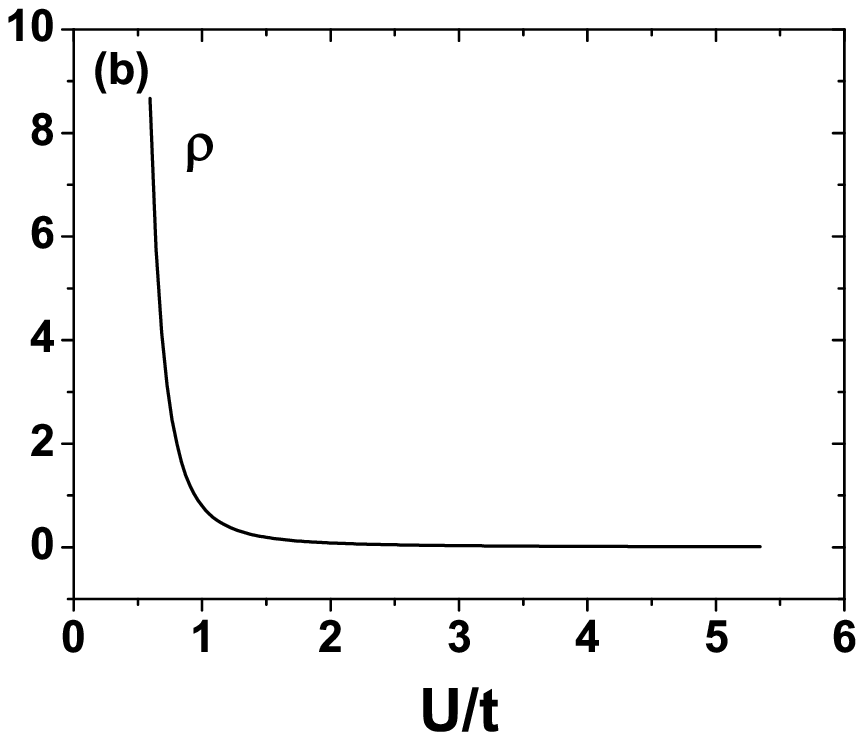}
\caption{Spin-quadrupole density wave velocity $v_{s}$ (solid line) compared
with the spin density wave velocity $v_{SDW}$ (dashed line) (a) and
stiffness $\protect\rho $ (b) in 2D for the staggered spin nematic phase. We
choose an $2U$ on-site interaction in the Hubbard model to obtain the same
gap equation as for our spin-quadrupole ordered phase.}
\end{figure}

Below $T\thicksim \Delta $, the single-particle excitations are gapped, and
the only contributions to the entropy are due to spin-quadrupole density
waves with a linear dispersion. On a 2D square lattice, the thermal energy
per unit volume is easily calculated as $E/T\approx 4.8(k_{B}T)^{3}/(\pi
\hbar ^{2}v_{s}^{2})$. When measuring the low temperature special heat, we
expect that $c\sim T^{2}$, from which $v_{s}$ can be determined. We would
like to mention that such a quadratic temperature dependence of the specific
heat is consistent with recent experiments on Ni$_{2}$Ga$_{2}$S$_{4}$ \cite%
{Nakatsuji-2005}, a rare example which, probably, has a spin-quadrupole
long-range ordered ground state \cite{Tsunetsugu-2005,Mila-2006,Senthil-2006}%
.

\section{Effective hyperfine spin-exchange interactions of the Mott states}

So far, a functional integral approach has been developed to
consider the generalized half filled spin-$\frac{3}{2}$ Hubbard
model, and we have obtained a number of interesting results. In
this Section, we derive the low-energy effective hyperfine spin
model Hamiltonian in the strong-coupling limit. In the Mott
insulating state with a small, but finite atomic tunneling
parameter, the virtual tunneling of ultracold atoms can induce
effective spin-exchange interactions between the nearest neighbor
sites, leading to possible magnetic multipolar ordering or
disordered states. In order to derive such interactions, we have
to carefully consider the strong coupling limit and perform the
second-order perturbation calculation.

\subsection{Single lattice site}

Let us first look at the extreme limit of $t=0$. Each lattice site decouples
from its nearest neighbor sites, and the single site Hamiltonian is obtained%
\begin{equation}
H_{0}=\frac{c_{0}}{2}N(N-1)+\frac{c_{2}}{2}(\mathbf{S}^{2}-\frac{15}{4}N).
\end{equation}%
The corresponding energy eigenstates labelled by $\left\vert
n,S,m\right\rangle $ with good quantum numbers $n$, $S$ and its z component $%
m$ are given by%
\begin{equation}
E_{0}=\frac{c_{0}}{2}n(n-1)+\frac{c_{2}}{2}\left( S(S+1)-\frac{15}{4}%
n\right) ,
\end{equation}%
which has $2S+1$-fold degeneracy. According to the Pauli principle, we can
construct the eigenstates with zero and even number of particles as%
\begin{eqnarray}
\left\vert 0,0,0\right\rangle &=&\left\vert \Omega \right\rangle ;  \notag \\
\left\vert 2,0,0\right\rangle &=&\frac{1}{\sqrt{2}}(\psi _{\frac{3}{2}%
}^{\dag }\psi _{-\frac{3}{2}}^{\dag }-\psi _{\frac{1}{2}}^{\dag }\psi _{-%
\frac{1}{2}}^{\dag })\left\vert \Omega \right\rangle ,  \notag \\
\left\vert 2,2,0\right\rangle &=&\frac{1}{\sqrt{2}}(\psi _{\frac{3}{2}%
}^{\dag }\psi _{-\frac{3}{2}}^{\dag }+\psi _{\frac{1}{2}}^{\dag }\psi _{-%
\frac{1}{2}}^{\dag })\left\vert \Omega \right\rangle ,  \notag \\
\left\vert 2,2,2\right\rangle &=&\psi _{\frac{3}{2}}^{\dag }\psi _{\frac{1}{2%
}}^{\dag }\left\vert \Omega \right\rangle ,\text{ \ }\left\vert
2,2,-2\right\rangle =\psi _{-\frac{1}{2}}^{\dag }\psi _{-\frac{3}{2}}^{\dag
}\left\vert \Omega \right\rangle ,  \notag \\
\left\vert 2,2,1\right\rangle &=&\psi _{\frac{3}{2}}^{\dag }\psi _{-\frac{1}{%
2}}^{\dag }\left\vert \Omega \right\rangle ,\text{ }\left\vert
2,2,-1\right\rangle =\psi _{\frac{1}{2}}^{\dag }\psi _{-\frac{3}{2}}^{\dag
}\left\vert \Omega \right\rangle ;  \notag \\
\left\vert 4,0,0\right\rangle &=&\psi _{\frac{3}{2}}^{\dag }\psi _{\frac{1}{2%
}}^{\dag }\psi _{-\frac{1}{2}}^{\dag }\psi _{-\frac{3}{2}}^{\dag }\left\vert
\Omega \right\rangle ,
\end{eqnarray}%
while the eigenstates with odd number of particles are given by%
\begin{eqnarray}
\left\vert 1,\frac{3}{2},\frac{3}{2}\right\rangle &=&\psi _{\frac{3}{2}%
}^{\dag }\left\vert \Omega \right\rangle ,\text{ \ }\left\vert 1,\frac{3}{2}%
,-\frac{3}{2}\right\rangle =\psi _{-\frac{3}{2}}^{\dag }\left\vert \Omega
\right\rangle ,  \notag \\
\left\vert 1,\frac{3}{2},\frac{1}{2}\right\rangle &=&\psi _{\frac{1}{2}%
}^{\dag }\left\vert \Omega \right\rangle ,\text{ \ }\left\vert 1,\frac{3}{2}%
,-\frac{1}{2}\right\rangle =\psi _{-\frac{1}{2}}^{\dag }\left\vert \Omega
\right\rangle ;  \notag \\
\left\vert 3,\frac{3}{2},\frac{3}{2}\right\rangle &=&\psi _{\frac{3}{2}%
}^{\dag }\psi _{\frac{1}{2}}^{\dag }\psi _{-\frac{1}{2}}^{\dag }\left\vert
\Omega \right\rangle ,  \notag \\
\left\vert 3,\frac{3}{2},-\frac{3}{2}\right\rangle &=&\psi _{\frac{1}{2}%
}^{\dag }\psi _{-\frac{1}{2}}^{\dag }\psi _{-\frac{3}{2}}^{\dag }\left\vert
\Omega \right\rangle ,  \notag \\
\left\vert 3,\frac{3}{2},\frac{1}{2}\right\rangle &=&\psi _{\frac{3}{2}%
}^{\dag }\psi _{\frac{1}{2}}^{\dag }\psi _{-\frac{3}{2}}^{\dag }\left\vert
\Omega \right\rangle ,  \notag \\
\left\vert 3,\frac{3}{2},-\frac{1}{2}\right\rangle &=&\psi _{\frac{3}{2}%
}^{\dag }\psi _{-\frac{1}{2}}^{\dag }\psi _{-\frac{3}{2}}^{\dag }\left\vert
\Omega \right\rangle .
\end{eqnarray}%
Here it is worth noticing that \textit{the number of the possible
eigenstates are substantially reduced by the Pauli exclusive principle}. The
corresponding eigen-energies are listed in Table (\ref{tab:table1}).
\begin{table}[tbp]
\caption{Eigenvalues of the single-site model Hamiltonian.}
\label{tab:table1}%
\begin{ruledtabular}
\begin{tabular}{lcr}
Particle number $n$ & Total spin $S$ & Energy $E_{0}$\\
\hline
0 & 0 & 0\\
1 & 3/2 & 0\\
2 & 0 & $c_{0}-\frac{15}{4}c_{2}$\\
2 & 2 & $c_{0}-\frac{3}{4}c_{2}$\\
3 & 3/2 & $3c_{0}-\frac{15}{4}c_{2}$\\
4 & 0 & $6c_{0}-\frac{15}{2}c_{2}$\\
\end{tabular}
\end{ruledtabular}
\end{table}

From the experimental point of view, the Mott insulating states with one or
two particles per site are the most interesting cases because they are free
of three-body decays.

\subsection{One atom per site}

To derive the effective model Hamiltonian, we can simply consider a two-site
problem. For one atom per site, the unperturbed ground state will be given
by $\left\vert 1,3/2,m_{i}\right\rangle ^{(i)}\left\vert
1,3/2,m_{j}\right\rangle ^{(j)}$. Then, four intermediate states are allowed:%
\begin{eqnarray}
\left\vert \Omega \right\rangle ^{(i)}\left\vert 2,2,m\right\rangle ^{(j)}%
\text{ and }i &\leftrightarrow &j,  \notag \\
\left\vert \Omega \right\rangle ^{(i)}\left\vert 2,0,0\right\rangle ^{(j)}%
\text{ and }i &\leftrightarrow &j.
\end{eqnarray}%
Since the tunneling processes of $H_{t}=-t\sum_{\alpha }(\psi _{i\alpha
}^{\dag }\psi _{j\alpha }+$H.c.$)$ conserve the total spin $\mathbf{S=S}_{i}%
\mathbf{+S}_{j}$, the total magnetic quantum number $m=m_{i}+m_{j}$, and the
total number of particles $n=n_{i}+n_{j}$ \cite{Demler-2003,Zawitkowski-2006}%
, the energy shifts of the total spin-$S$ channel from the second-order
perturbation theory can be calculated as%
\begin{equation}
\epsilon _{S}=-t^{2}\sum_{\upsilon }\frac{|\left\langle \upsilon \right\vert
H_{t}\left\vert n,S,m\right\rangle |^{2}}{E_{\upsilon }-E_{g}},
\end{equation}%
where $\left\vert n,S,m\right\rangle $ and $|\upsilon \rangle $ denote the
initial state and the possible intermediate states in the representation of
the two-site system, respectively. The total number of particles $n$, the
total spin $\mathbf{S}$, and its z component $S^{z}$ are good quantum
numbers, while $E_{\upsilon }$ and $E_{g}$ correspond to the zeroth-order
eigenenergies of these states, which are calculated from the single-site
eigenenergies in the Table I. According to the allowed intermediate states,
the channels with total spin $S=0$ and $S=2$ have to be considered in this
case.

For the energy shift in the total spin $S=0$ channel, there are two possible
intermediate states: $S_{i}=0,$ $n_{i}=0;$ $S_{j}=0,$ $n_{j}=2$ and $%
i\leftrightarrow j$. With the help of the Clebsch-Gordon coefficients, the
unperturbed state and intermediate states are expressed in the form%
\begin{eqnarray}
\left| 2,0,0\right\rangle &=&\frac{1}{2}\left( \left| 1,\frac{3}{2},\frac{3}{%
2}\right\rangle ^{(i)}\left| 1,\frac{3}{2},-\frac{3}{2}\right\rangle
^{(j)}\right.  \notag \\
&&-\left| 1,\frac{3}{2},-\frac{3}{2}\right\rangle ^{(i)}\left| 1,\frac{3}{2},%
\frac{3}{2}\right\rangle ^{(j)}  \notag \\
&&-\left| 1,\frac{3}{2},\frac{1}{2}\right\rangle ^{(i)}\left| 1,\frac{3}{2},-%
\frac{1}{2}\right\rangle ^{(j)}  \notag \\
&&\left. +\left| 1,\frac{3}{2},-\frac{1}{2}\right\rangle ^{(i)}\left| 1,%
\frac{3}{2},\frac{1}{2}\right\rangle ^{(j)}\right) ,  \notag \\
\left| 2,0,0\right\rangle _{\text{int}} &=&\left| \Omega \right\rangle
^{(i)}\left| 2,0,0\right\rangle ^{(j)},\text{ and }i\leftrightarrow j.
\end{eqnarray}
Thus the corresponding energy shift is evaluated as%
\begin{equation}
\epsilon _{0}=-\frac{16t^{2}}{4c_{0}-15c_{2}}.
\end{equation}

Then let us consider the total spin $S=2$ channel. There are two possible
intermediate states $S_{i}=0,$ $n_{i}=0,$ $S_{j}=2,$ $n_{j}=2$ and $%
i\leftrightarrow j$. We notice that evaluation of the energy shift with
maximal spin projection $m=S$ is sufficient for obtaining the needed
results, because the tunneling Hamiltonian is SU(2) spin invariant and the
tunneling processes do not mix the different $m_{F}$ states. In the
representation of the two-site system with good quantum numbers, the
unperturbed state and intermediate states with maximal spin polarization are
written as%
\begin{eqnarray}
&&\left| 2,2,2\right\rangle =\frac{1}{\sqrt{2}}\left| 1,\frac{3}{2},\frac{3}{%
2}\right\rangle ^{(i)}\left| 1,\frac{3}{2},\frac{1}{2}\right\rangle ^{(j)}
\notag \\
&&\text{ \ \ \ \ \ \ \ \ \ \ }-\frac{1}{\sqrt{2}}\left| 1,\frac{3}{2},\frac{1%
}{2}\right\rangle ^{(i)}\left| 1,\frac{3}{2},\frac{3}{2}\right\rangle ^{(j)},
\notag \\
&&\left| 2,2,2\right\rangle _{\text{int}}=\left| \Omega \right\rangle
^{(i)}\left| 2,2,2\right\rangle ^{(j)},\text{ and }i\leftrightarrow j.
\end{eqnarray}%
Similar calculation leads to the energy shift in the total spin $S=2$
channel as%
\begin{equation}
\epsilon _{2}=-\frac{16t^{2}}{4c_{0}-3c_{2}}.
\end{equation}

Thus, up to the second order of the hopping term, the effective
spin-exchange interactions for the two-site problem are obtained as%
\begin{equation}
H_{ij}=\epsilon _{0}\mathcal{P}_{ij}(0)+\epsilon _{2}\mathcal{P}_{ij}(2),
\end{equation}%
where $\mathcal{P}_{ij}(S)$ projects the two-spin states of $S_{i}=S_{j}=3/2$
onto the total spin-$S$ state. The explicit form of $\mathcal{P}_{ij}(S)$
are given as follows:%
\begin{eqnarray}
\mathcal{P}_{ij}(0) &=&\frac{\left( \mathbf{S}_{i}\mathbf{\cdot S}%
_{j}-\lambda _{1}\right) \left( \mathbf{S}_{i}\mathbf{\cdot S}_{j}-\lambda
_{2}\right) \left( \mathbf{S}_{i}\mathbf{\cdot S}_{j}-\lambda _{3}\right) }{%
\left( \lambda _{0}-\lambda _{1}\right) \left( \lambda _{0}-\lambda
_{2}\right) \left( \lambda _{0}-\lambda _{3}\right) },  \notag \\
\mathcal{P}_{ij}(2) &=&\frac{\left( \mathbf{S}_{i}\mathbf{\cdot S}%
_{j}-\lambda _{0}\right) \left( \mathbf{S}_{i}\mathbf{\cdot S}_{j}-\lambda
_{1}\right) \left( \mathbf{S}_{i}\mathbf{\cdot S}_{j}-\lambda _{3}\right) }{%
\left( \lambda _{2}-\lambda _{0}\right) \left( \lambda _{2}-\lambda
_{1}\right) \left( \lambda _{2}-\lambda _{3}\right) },  \label{projector}
\end{eqnarray}%
with $\lambda _{S}=\frac{1}{2}\left[ S(S+1)-\frac{15}{2}\right] $. For the
lattice model, the effective spin-exchange interactions is finally obtained
up to a constant
\begin{equation}
H_{eff}=\epsilon _{0}\sum_{<ij>}\mathcal{P}_{ij}(0)+\epsilon _{2}\sum_{<ij>}%
\mathcal{P}_{ij}(2).
\end{equation}%
Actually, a similar effective hyperfine spin model Hamiltonian was also
given in the earlier work \cite{YPWang-2005}. Here we would emphasize that
the validity of the second-order perturbation theory is restricted to $%
c_{0}\gg t$ to ensure the stability of the Mott insulating state. When $t$
becomes large and comparable to the energy difference between unperturbed
ground state and the lowest intermediate states, higher-order perturbations
should be included, and the spin exchange beyond the nearest neighbor sites
should be considered.

In particular, when $c_{2}=0$, the original spin-$\frac{3}{2}$ Hubbard model
displays an SU(4) symmetry \cite{CWu-2003,Hofstetter-2004}, and here we have
$\epsilon _{0}=\epsilon _{2}=-\frac{4t^{2}}{c_{0}}$. Then the effective
spin-exchange interaction becomes%
\begin{equation}
H_{eff}=-\frac{4t^{2}}{c_{0}}\sum_{<ij>}\left[ \mathcal{P}_{ij}(0)+\mathcal{P%
}_{ij}(2)\right] ,
\end{equation}%
with%
\begin{eqnarray}
&&\mathcal{P}_{ij}(0)+\mathcal{P}_{ij}(2)  \notag \\
&=&-\frac{1}{9}\left( \mathbf{S}_{i}\mathbf{\cdot S}_{j}\right) ^{3}-\frac{11%
}{36}\left( \mathbf{S}_{i}\mathbf{\cdot S}_{j}\right) ^{2}+\frac{9}{16}%
\mathbf{S}_{i}\mathbf{\cdot S}_{j}+\frac{99}{64}.
\end{eqnarray}%
This effective spin model also exhibits a uniform SU(4) symmetry.
In one dimension, an
exact solution had been obtained by the Bethe ansatz method \cite%
{Sutherland-1975}, and the ground state of the SU(4) spin-exchange model is
a spin singlet with gapless spin excitations. Moreover, such a spin-$\frac{3%
}{2}$ exchange Hamiltonian is equivalent to the so-called SU(4) spin-orbital
model, which has been extensively studied in solid-state systems \cite%
{spin-orbit}.

\subsection{Two atoms per site}

More interesting and complicated situations occur for a Mott insulator with
two interacting atoms per site. Now the possible ground states on each site
may depend on the spin dependent coupling parameter $c_{2}$. In the limit of
$t=0$, the single site state is a spin singlet ($S_{i}=0$) for $c_{2}>0$ and
a spin quintet ($S_{i}=2$) for $c_{2}<0$. However, a small finite tunneling
induces an exchange energy of order $t^{2}/c_{0}$ between the nearest
neighbor sites. Since the energy difference between the spin singlet and
quintet states is given by $c_{2}$, when $t^{2}/c_{0}$ is large compared
with the absolute value of $c_{2}$, the quintet state is mixed with the
singlet state, and both spin configurations of these two states have to be
taken into account in the second-order perturbation calculations. However,
in the present section we will not consider this situation and simply assume
$t\ll \sqrt{c_{0}|c_{2}|}$. The discussion is divided into two parts.

\subsubsection{$c_{2}>0$}

In this case, the two-site unperturbed ground state corresponds to the site
singlet state $\left| 2,0,0\right\rangle ^{(i)}\left| 2,0,0\right\rangle
^{(j)}$, and the allowed intermediate states are%
\begin{equation}
\left| 1,\frac{3}{2},m_{i}\right\rangle ^{(i)}\left| 3,\frac{3}{2}%
,m_{j}\right\rangle ^{(j)}\text{ and }i\leftrightarrow j.
\label{intermediate}
\end{equation}%
According to the initial state, only the total spin $S=0$ channel is
involved. The two-site unperturbed and intermediate states are expressed as%
\begin{eqnarray}
\left| 4,0,0\right\rangle &=&\left| 2,0,0\right\rangle ^{(i)}\left|
2,0,0\right\rangle ^{(j)},  \notag \\
\left| 4,0,0\right\rangle _{\text{int}} &=&\frac{1}{2}\left| 1,\frac{3}{2},%
\frac{3}{2}\right\rangle ^{(i)}\left| 3,\frac{3}{2},-\frac{3}{2}%
\right\rangle ^{(j)}  \notag \\
&&-\frac{1}{2}\left| 1,\frac{3}{2},-\frac{3}{2}\right\rangle ^{(i)}\left| 3,%
\frac{3}{2},\frac{3}{2}\right\rangle ^{(j)}  \notag \\
\text{ \ \ \ \ } &&-\frac{1}{2}\left| 1,\frac{3}{2},\frac{1}{2}\right\rangle
^{(i)}\left| 3,\frac{3}{2},-\frac{1}{2}\right\rangle ^{(j)}  \notag \\
&&+\frac{1}{2}\left| 1,\frac{3}{2},-\frac{1}{2}\right\rangle ^{(i)}\left| 3,%
\frac{3}{2},\frac{1}{2}\right\rangle ^{(j)},  \label{spin0}
\end{eqnarray}%
and the corresponding intermediate state with $i\leftrightarrow j$. After
some algebra, the energy shift is calculated as%
\begin{equation}
\epsilon _{0}=-\frac{8t^{2}}{4c_{0}+15c_{2}}.
\end{equation}%
Then, up to a constant, the effective lattice model is obtained as%
\begin{equation}
H_{eff}=-\frac{8t^{2}}{4c_{0}+15c_{2}}\sum_{<ij>}\mathcal{P}_{ij}(0),
\end{equation}%
where $\mathcal{P}_{ij}(0)$ is the singlet projection operator between two
site singlet states, which can not be expressed in terms of powers of
hyperfine spin exchanges. In this sense, the functional integral approach
developed in Sec. III is a more appropriate method to attack this problem.

\subsubsection{$c_{2}<0$}

In this case the two-site unperturbed ground state is described by the
quintet state $\left| 2,2,m_{i}\right\rangle ^{(i)}\left|
2,2,m_{j}\right\rangle ^{(j)}$. Two intermediate states have been given in (%
\ref{intermediate}), but the total spin $S=0,1,2,3$ channels have to be
considered, as the virtual tunneling process is forbidden in the $S=4$
channel due to the Pauli's exclusion principle. With the help of the
Clebsch-Gordon coefficients, the unperturbed two-site ground states with
maximal spin polarization are written as%
\begin{eqnarray}
\left| 4,3,3\right\rangle &=&\frac{1}{\sqrt{2}}\left| 2,2,2\right\rangle
^{(i)}\left| 2,2,1\right\rangle ^{(j)}  \notag \\
&&-\frac{1}{\sqrt{2}}\left| 2,2,1\right\rangle ^{(i)}\left|
2,2,2\right\rangle ^{(j)},  \notag \\
\left| 4,2,2\right\rangle &=&\sqrt{\frac{2}{7}}\left| 2,2,2\right\rangle
^{(i)}\left| 2,2,0\right\rangle ^{(j)}  \notag \\
&&+\sqrt{\frac{2}{7}}\left| 2,2,0\right\rangle ^{(i)}\left|
2,2,2\right\rangle ^{(j)}  \notag \\
&&-\sqrt{\frac{3}{7}}\left| 2,2,1\right\rangle ^{(i)}\left|
2,2,1\right\rangle ^{(j)},  \notag \\
\left| 4,1,1\right\rangle &=&\frac{1}{\sqrt{5}}\left| 2,2,2\right\rangle
^{(i)}\left| 2,2,-1\right\rangle ^{(j)}  \notag \\
&&-\frac{1}{\sqrt{5}}\left| 2,2,-1\right\rangle ^{(i)}\left|
2,2,2\right\rangle ^{(j)}  \notag \\
&&-\sqrt{\frac{3}{10}}\left| 2,2,1\right\rangle ^{(i)}\left|
2,2,0\right\rangle ^{(j)}  \notag \\
&&+\sqrt{\frac{3}{10}}\left| 2,2,0\right\rangle ^{(i)}\left|
2,2,1\right\rangle ^{(j)},  \notag \\
\left| 4,0,0\right\rangle &=&\frac{1}{\sqrt{5}}\left| 2,2,2\right\rangle
^{(i)}\left| 2,2,-2\right\rangle ^{(j)}  \notag \\
&&+\frac{1}{\sqrt{5}}\left| 2,2,0\right\rangle ^{(i)}\left|
2,2,0\right\rangle ^{(j)}  \notag \\
&&-\frac{1}{\sqrt{5}}\left| 2,2,-1\right\rangle ^{(i)}\left|
2,2,1\right\rangle ^{(j)}  \notag \\
&&-\frac{1}{\sqrt{5}}\left| 2,2,1\right\rangle ^{(i)}\left|
2,2,-1\right\rangle ^{(j)}  \notag \\
&&+\frac{1}{\sqrt{5}}\left| 2,2,-2\right\rangle ^{(i)}\left|
2,2,2\right\rangle ^{(j)}.
\end{eqnarray}%
\newline
The intermediate states in total spin $S=0$ channel have been expressed as (%
\ref{spin0}), while the other intermediate states in channel $S=3,2,1$ are
expressed as%
\begin{eqnarray}
\left| 4,3,3\right\rangle _{\text{int}} &=&\left| 1,\frac{3}{2},\frac{3}{2}%
\right\rangle ^{(i)}\left| 3,\frac{3}{2},\frac{3}{2}\right\rangle ^{(j)},
\notag \\
\left| 4,2,2\right\rangle _{\text{int}} &=&\frac{1}{\sqrt{2}}\left| 1,\frac{3%
}{2},\frac{3}{2}\right\rangle ^{(i)}\left| 3,\frac{3}{2},\frac{1}{2}%
\right\rangle ^{(j)}  \notag \\
&&-\frac{1}{\sqrt{2}}\left| 1,\frac{3}{2},\frac{1}{2}\right\rangle
^{(i)}\left| 3,\frac{3}{2},\frac{3}{2}\right\rangle ^{(j)},  \notag \\
\left| 4,1,1\right\rangle _{\text{int}} &=&\sqrt{\frac{3}{10}}\left| 1,\frac{%
3}{2},\frac{3}{2}\right\rangle ^{(i)}\left| 3,\frac{3}{2},-\frac{1}{2}%
\right\rangle ^{(j)}  \notag \\
&&-\sqrt{\frac{2}{5}}\left| 1,\frac{3}{2},\frac{1}{2}\right\rangle
^{(i)}\left| 3,\frac{3}{2},\frac{1}{2}\right\rangle ^{(j)}  \notag \\
&&+\sqrt{\frac{3}{10}}\left| 1,\frac{3}{2},-\frac{1}{2}\right\rangle
^{(i)}\left| 3,\frac{3}{2},\frac{3}{2}\right\rangle ^{(j)},
\end{eqnarray}%
and corresponding intermediate states with interchange $i\leftrightarrow j$.
It is sufficient to calculate the energy shifts with these maximal spin
polarized states. The final results are obtained as%
\begin{equation}
\epsilon _{0}=-\frac{40t^{2}}{4c_{0}-9c_{2}},\text{ }\epsilon _{1}=\epsilon
_{3}=\frac{2}{5}\epsilon _{0},\text{ }\epsilon _{2}=0.
\end{equation}%
Note that up to the second order of $t$, the total spin $S=2$ channel makes
no contribution to the energy shift. Hence, the resulting effective spin
interaction for the lattice model reads up to a constant%
\begin{equation}
H_{eff}=\epsilon _{0}\sum_{<ij>}\left\{ \mathcal{P}_{ij}(0)+\frac{2}{5}\left[
\mathcal{P}_{ij}(1)+\mathcal{P}_{ij}(3)\right] \right\} ,
\end{equation}%
where $\mathcal{P}_{ij}(S)$ projects the $S_{i}=S_{j}=2$ states onto the
total spin-$S$ states. They are given by
\begin{widetext}
\begin{eqnarray}
\mathcal{P}_{ij}(0) &=&\frac{\left( \mathbf{S}_{i}\mathbf{\cdot S}%
_{j}-\lambda _{1}\right) \left( \mathbf{S}_{i}\mathbf{\cdot S}_{j}-\lambda
_{2}\right) \left( \mathbf{S}_{i}\mathbf{\cdot S}_{j}-\lambda _{3}\right)
\left( \mathbf{S}_{i}\mathbf{\cdot S}_{j}-\lambda _{4}\right) }{\left(
\lambda _{0}-\lambda _{1}\right) \left( \lambda _{0}-\lambda _{2}\right)
\left( \lambda _{0}-\lambda _{3}\right) \left( \lambda _{0}-\lambda
_{4}\right) },  \notag \\
\mathcal{P}_{ij}(1) &=&\frac{\left( \mathbf{S}_{i}\mathbf{\cdot S}%
_{j}-\lambda _{0}\right) \left( \mathbf{S}_{i}\mathbf{\cdot S}_{j}-\lambda
_{2}\right) \left( \mathbf{S}_{i}\mathbf{\cdot S}_{j}-\lambda _{3}\right)
\left( \mathbf{S}_{i}\mathbf{\cdot S}_{j}-\lambda _{4}\right) }{\left(
\lambda _{1}-\lambda _{0}\right) \left( \lambda _{1}-\lambda _{2}\right)
\left( \lambda _{1}-\lambda _{3}\right) \left( \lambda _{1}-\lambda
_{4}\right) },  \notag \\
\mathcal{P}_{ij}(3) &=&\frac{\left( \mathbf{S}_{i}\mathbf{\cdot S}%
_{j}-\lambda _{0}\right) \left( \mathbf{S}_{i}\mathbf{\cdot S}_{j}-\lambda
_{1}\right) \left( \mathbf{S}_{i}\mathbf{\cdot S}_{j}-\lambda _{2}\right)
\left( \mathbf{S}_{i}\mathbf{\cdot S}_{j}-\lambda _{4}\right) }{\left(
\lambda _{3}-\lambda _{0}\right) \left( \lambda _{3}-\lambda _{1}\right)
\left( \lambda _{3}-\lambda _{2}\right) \left( \lambda _{3}-\lambda
_{4}\right) },
\end{eqnarray}
\end{widetext}with $\lambda _{S}=\frac{1}{2}\left[ S(S+1)-12\right] $.
Moreover, the effective spin-exchange model Hamiltonian can be
simply
written as%
\begin{equation}
H_{eff}=-\frac{\epsilon _{0}}{6}\sum_{<ij>}\left[ \frac{1}{15}\left( \mathbf{%
S}_{i}\mathbf{\cdot S}_{j}\right) ^{3}+\frac{2}{15}\left( \mathbf{S}_{i}%
\mathbf{\cdot S}_{j}\right) ^{2}-\mathbf{S}_{i}\mathbf{\cdot S}_{j}\right] .
\end{equation}%
Here we would like to emphasize that the spin operators correspond to $S=2$
operators.

\section{Concluding remarks}

To summarize, a functional integral approach has been applied to
study the quantum spin-quadrupole long-range ordered state of the
Mott insulating
phase in the generalized half filled spin-$\frac{3}{2}$ Hubbard model for the case $%
a_{2}>a_{0}>0$. On a square lattice, the ground state shows a
staggered spin-quadrupole long-range order from the saddle-point
solution of the effective action. By including the Gaussian
fluctuations, the four gapless collective modes have been found,
corresponding to the spin-quadrupole density waves. Unlike spin
density waves in the half filled spin-$\frac{1}{2} $ Hubbard
model, the spin-quadrupole density wave velocity is saturated in
the limit of $a_{2}\gg a_{0}$, because the transverse mode
couplings of the spin-quadrupole collective excitations are not
allowed by the SO(4) invariant effective action. Thus, our results
obtained under the Gaussian approximation for the spin-quadrupole
ordered state are robust even when the high order fluctuations are
included. This is a remarkable property of the enlarged hyperfine
spin space dimensionality and the higher symmetry of the local
interactions in the generalized spin-$\frac{3}{2}$ Hubbard model.

Moreover, the effective hyperfine spin-exchange interactions for
the quarter filled and half filled cases have been derived from
the second-order perturbation theory. Starting from these
effective spin-exchange interactions, more efficient treatments
may be carried out to characterize these multipolar magnetic
states. Further investigations in this regard will be reported
elsewhere.

G.-M.Z. would like to thank Tao Xiang for his critical comments
and constructive suggestions. We acknowledge the support of
NSF-China (No.10125418 and 10474051).

\end{document}